\documentclass[aps, prd,10pt,notitlepage,nofootinbib,superscriptaddress,showkeys]{revtex4-1}
\usepackage{amstext,amsmath,amssymb,amsfonts,bbm}
\usepackage[latin1]{inputenc}
\usepackage{graphicx}
\usepackage{hyperref}
\usepackage{epstopdf}
\usepackage{amsthm}
\usepackage{pstricks}
\usepackage{tocvsec2}

\usepackage{color}

\def\beq{\begin{equation}}
\def\be{\begin{equation}}
\def\ee{\end{equation}}
\def\bes{\begin{eqnarray}}
\def\ees{\end{eqnarray}}

\DeclareMathOperator{\Ad}{Ad}

\DeclareMathOperator{\SU}{SU}
\DeclareMathOperator{\SO}{SO}

\newcommand{\su}{\mathfrak{su}}

\newcommand{\unit}{\mathbbm{1}}

\def\f{\frac}

\def\mone{^{-1}}

\def\N{{\mathbbm N}}

\def\R{{\mathbbm R}}

\def\calH{{\mathcal H}}

\theoremstyle{definition}
\theoremstyle{definition}
\theoremstyle{definition}
\theoremstyle{definition}
\theoremstyle{definition}
\theoremstyle{definition}

\begin{document}
\maxtocdepth{subsection}

\title{\large \bf A taste of Hamiltonian constraint in spin foam models}

\author{{\bf Valentin Bonzom}}\email{vbonzom@perimeterinstitute.ca}

\affiliation{Perimeter Institute for Theoretical Physics, 31 Caroline St. N, ON N2L 2Y5, Waterloo, Canada}

\date{\small\today}

\begin{abstract}\noindent
The asymptotics of some spin foam amplitudes for a quantum 4-simplex is known to display rapid oscillations whose frequency is the Regge action. In this note, we reformulate this result through a difference equation, asymptotically satisfied by these models, and whose semi-classical solutions are precisely the sine and the cosine of the Regge action. This equation is then interpreted as coming from the canonical quantization of a simple constraint in Regge calculus. This suggests to lift and generalize this constraint to the phase space of loop quantum gravity parametrized by twisted geometries. The result is a reformulation of the flat model for topological BF theory from the Hamiltonian perspective. The Wheeler-de-Witt equation in the spin network basis gives difference equations which are exactly recursion relations on the 15j-symbol. Moreover, the semi-classical limit is investigated using coherent states, and produces the expected results. It mimics the classical constraint with quantized areas, and for Regge geometries it reduces to the semi-classical equation which has been introduced in the beginning.
\end{abstract}

\maketitle

\section*{Introduction}
\subsection*{Asymptotics of spin foam amplitudes from semi-classical Hamiltonian dynamics}

A good spin foam model for quantum gravity is (often) expected to reproduce Regge calculus (a large distance approximation of general relativity) in the classical limit. The idea goes back to Ponzano and Regge, \cite{PR}, who made the key observation that the Wigner 6j-symbol, an object from the theory of representations of $\SU(2)$, behaves for large spins as the cosine of the Regge action for a tetrahedron, with the spins as edge lengths. This gave a model for quantum gravity in three dimensions, where spins are interpreted as quantized lengths. The achievement of Loop quantum gravity (LQG) then gave a new birth and justification to the idea that quantum gravity can be formulated from algebraic objects, coming from the representation theory of a Lie group, attached to chunks of spacetime (simplices, polyhedra).

Still, it is not so obvious and straightforward to imagine why the semi-classical limit of spin foams would have to be expressed in terms of the Regge approximation to general relativity (remember spin foams are initially designed to provide transition amplitudes between the kinematical states of LQG, based on cylindrical functionals of the Ashtekar-Barbero connection, \cite{perez-intro-lqg}). This idea was suggested in \cite{rovelli-PRTVO}. In particular, it is based on the fact that LQG supports a discrete area spectrum (built from the Casimir of $\SU(2)$), very similar to the Ponzano-Regge ansatz for quantized lengths in three dimensions.

So when a new model is proposed, the natural thing that is to be done is to check its semi-classical limit, where by ``checking`` it is usually meant chasing after the Regge action. However, it turns out that the models that have so far attracted the most attention all have such Regge contributions, in particular models that are known {\it not} to describe quantum gravity, like the Ooguri model (a model for the topological BF theory in four dimensions, \cite{ooguri4d}) and the Barrett-Crane model (though the oscillatory part involving the Regge action is only a subleading term \cite{freidel-louapre-6j}).

In a series of papers (see for instance \cite{barrett-asym15j, barrett-asym-summary}), the asymptotics of the 15j-symbol (for the Ooguri model) and for the Euclidean and Lorentzian EPRL 4-simplex have been precisely studied. It appears that different behaviours are observed according to whether or not the boundary data determine a Regge metric on the 4-simplex, so that oscillations with the Regge action do occur or do not. 

So we would like to get a {\it criterion} which would tell us whenever a model has such oscillations with the Regge action, say at the leading order. We obtain such a criterion as a difference equation of second order on the 4-simplex amplitude, \eqref{semiclass eq}. This equation is actually well-known from the three-dimensional case. Indeed the 6j-symbol is fully characterized by a second order recursion relation (coming from the Biendenharn-Elliott, or pentagon identity), which, although generally complicated, simplifies in the semi-classical limit where it allows to determine the asymptotics, \cite{schulten-gordon2, maite-etera-6jcorr}.

Since the asymtotics provides a regime where the spin foam amplitude may be approximated by some quantum Regge calculus, it is natural to look for an interpretation, or even better, for a derivation of this difference equation as the quantization of a constraint in Regge calculus. It turned out to be very simple, and natural. The corresponding constraint is a sort of flatness constraint which enables to built a flat 4-simplex from its boundary. It states that the momenta conjugated to the triangle areas have to be the dihedral angles between adjacent tetrahedra, computed from the areas like in a flat 4-simplex. This gives a first link between a classical constraint and the asymptotics of spin foams.

So far the relation between spin foam models and the Hamiltonian constraint of general relativity has been particularly evasive (see open problem (14) in \cite{rovelli-new-look-lqg}). Our result gives a taster for this relation. To go further and ultimately savour it, we need to derive the semi-classical constraint from the quantization of a Hamiltonian operator in LQG. We will perform this task in the case of the Ooguri model. The analysis is a simple extension of results to appear from a collaboration with L. Freidel, \cite{3d-wdw}. There it is shown in the 3d case that a projection of the curvature onto the components of the triad, thus taking the form of the Hamiltonian constraint $EEF$, can be quantized in LQG. In the simplest situation, on the boundary of a tetrahedron, the Wheeler-de-Witt equation is a difference equation which is {\it exactly} the recursion relation defining the 6j-symbol. In 4d, the physical (flat) state on the boundary of a 4-simplex is the 15j-symbol. One can lift the Hamiltonian used in 3d to 4d, and using the methods of \cite{3d-wdw}, we claim that the Wheeler-de-Witt equation reproduces the recursion relations satisfied by the 15j-symbol which were derived in \cite{recurrence-paper}.

The organization is as follows. In the section \ref{sec:semiclass}:
\begin{itemize}
 \item we exhibit a difference equation whose solutions are the exponentials of $\pm i$ times the Regge action of the 4-simplex.
 \item we derive this equation as the quantization of a classical constraint in area Regge calculus. The constraint states that a point on the phase space is given by a set of ten areas and ten dihedral angles which are those of a flat 4-simplex, determined by the areas.
\end{itemize}
The section \ref{sec:4d-wdw} focuses on the Wheeler-de-Witt equation for the Ooguri model.
\begin{itemize}
 \item We define a classical constraint, attached to a node and a cycle of a spin network graph, by projecting the curvature onto some components of the gravitational field, \ref{sec:proposal}.
 \item This Hamiltonian is rewritten in terms of twisted geometries \cite{freidel-speziale-twisted-geom}, a nice parametrization of the LQG phase space on a single graph. It appears as a generalization of the above constraint for Regge calculus to the whole LQG phase space (including non-Regge boundary data), \ref{sec:twisted H}.
 \item The corresponding Wheeler-de-Witt equation is studied, more particularly in the large spin limit, in the coherent state basis. Using the WKB approximation, it reduces to the classical Hamiltonian on twisted geometries, with quantized areas.
\end{itemize}
In particular, in the Regge sector of boundary data, it reproduces the semi-classical equation asymptotically satisfied by spin foam models. The natural variables are areas and normals of triangles. Thus, it strengthens from the Hamiltonian point of view the result that quantum area-angle calculus is the semi-classical limit of quantized geometries (in the Regge sector).

We will also argue that such difference equations obtained through canonical quantization in the LQG framework lead to a position where the same analysis as that of \cite{barrett-asym15j} can be done and used to extract the asymptotics (like in 3d actually, \cite{3d-wdw,schulten-gordon2}).

In the section \ref{sec:outlook}, we discuss additional interesting difference equations. One is derived from our main semi-classical equation \eqref{semiclass eq}, and shown to probe the closure of the simplex (it was already introduced and precisely described in \cite{recurrence-paper}, though from a quite different path). We also sketch the possibility of introducing more speculative constraints, whose asymptotical behaviour exhibits oscillations with the Regge action.

All technical details are skipped in the main text to ease a fluent reading, and are reported in appendix.


\section{A new look at the semi-classical behaviour of 4-simplex spin foam amplitudes} \label{sec:semiclass}

Crucial references on the asymptotics of spin foam models are \cite{pereira-asymEPR, barrett-asymEPR, barrett-asym15j, barrett-asym-summary}. There it is shown that several spin foam models get in the large area limit rapid oscillations with the Regge action. We want to track back this phenomenon to the fact that they satisfy in this regime the same equation, solved by exponentials of $i$ times the Regge action. 


Consider a 4-simplex, with five tetrahedra on its boundary labelled by $a=1,\dotsc, 5$. The semi-classical regime corresponds to large values of the quantum numbers of triangle areas $(j_{ab}\in \f\N2)_{a<b}$. A spin foam amplitude for the 4-simplex is determined by some boundary data, including these quantum areas. The additional boundary data are coherent states labelled by spinors in the Lorentzian EPRL model \cite{pereira-asymEPR}, and by points on the 2-sphere $(\vec{N}_{ab}\in S^2)_{a,b}$ for the $\SU(2)$ Ooguri model \footnote{There are also some phase ambiguities, which will play an important role in the next sections, but we do not need such details at this stage.}. The latter have been introduced in spin foam models to gain geometric control on the quantum amplitude, \cite{livine-coherentBF}. Indeed, there are some distinguished sets of boundary data which allow to construct the geometry of genuine flat 4-simplices in $\R^4$. There, the vectors $(\vec{N}_{ab})$ are the normals to the triangles with areas $j_{ab}$, and the five tetrahedra consistently glue to form a 4-simplex. This is the geometric sector of the model, consisting in all possible Regge geometries. In this sector, it has been shown in a series of paper (see \cite{barrett-asym15j} and other references therein) that for homogeneously scaled areas $(\lambda\,j_{ab})$, with $\lambda \gg 1$,
\beq \label{asymptotics}
W(\lambda j_{ab}) = N_+\,e^{i\lambda S_{\rm R}(j_{ab})} + N_-\,e^{-i\lambda S_{\rm R}(j_{ab})} + o(1),
\ee
up to a global scaling. The coefficients $N_\pm$ depend on the boundary data, but not on $\lambda$. The Regge action is as usual:
\beq \label{regge action}
S_{\rm R}(j_{ab}) = \sum_{a<b} A_{ab}\,\Theta_{ab},
\ee
where the areas are actually given by: $A_{ab}=j_{ab}$ in the Ooguri model and $A_{ab}=\gamma\,j_{ab}$ in the EPRL model, the global parameter $\gamma$ being known as the Immirzi parameter (this last point is in agreement with the area spectrum of canonical Loop Quantum Gravity). The angles $\Theta_{ab}$ are the dihedral angles determined by the geometry of the boundary data, which are then the angles between the boundary tetrahedra.

So the large spin regime provides a notion of semi-classical quantum gravity, where the spin foam amplitude is well approximated by some quantum Regge calculus. Still, because this happens for several models in spite of their fundamental differences, one would like to understand this behaviour in a somewhat universal way. This is what we do by exhibiting a difference equation asymptotically satisfied by these models, which directly leads to this semi-classical approximation. Then we derive this equation as a naive quantization of some flatness constraint in area Regge calculus. To go further along these lines, it is desirable to get an equation satisfied by the full amplitude, and not only in the semi-classical limit.

\subsection{Recursion relations in the asymptotics}

Our first point is that linear combinations of exponentials of $\pm i$ times the Regge action are the solutions to the following difference equation of the second order:
\beq \label{semiclass eq}
\Bigl[\Delta_{ab} +2\,\bigl(1-\cos \Theta_{ab}\bigr)\Bigr] V(A_{ab}) = 0,
\ee
when solved via the semi-classical approximation for large $\lambda$. Here $\Delta_{ab}$ is the discrete second derivative with respect to the area variable $A_{ab}$ (may it be $j_{ab}$ or $\gamma j_{ab}$): $\Delta f(x) = f(x+1)+ f(x-1) -2 f(x)$. An equivalent form (which will be that naturally coming out in the next sections) is obtained by defining some ladder operators which shift an area by $\pm 1$,
\beq \label{op +-}
\delta^+_{ab} V(A_{ab}) = V(A_{ab}+1),\qquad \delta^-_{ab} V(A_{ab}) = V(A_{ab}-1).
\ee
Then, the semi-classical equation becomes:
\beq
\Bigl[\f12\bigl(\delta^+_{ab} + \delta^-_{ab}\bigr) - \cos\Theta_{ab}\Bigr]\ V(A_{ab}) = 0.
\ee

\bigskip

We look for solving the equation {\it \`a la WKB}, when all spins are rescaled by $\lambda \gg 1$, and with the ansatz:
\beq
\psi(\lambda A_{ab}) = \Phi(A_{ab})\,e^{i S(\lambda A_{ab})}.
\ee
We assume $\Phi$ does not scale with $\lambda$, while $S$ scales linearly, so the idea is as usual: a slowly varying amplitude, with a rapidly oscillatory phase. To zeroth order,
\beq
\psi(\lambda A_{ab}\pm 1) \simeq \psi(\lambda A_{ab})\,e^{\pm iS'(\lambda A_{ab})},
\ee
where $S'$ is the derivative of $S$ seen as a function on the real line. The equation \eqref{semiclass eq} becomes:
\beq \label{solve semiclass}
\cos S'(A_{ab}) - \cos\Theta_{ab} = 0,
\ee
or: $S'(A_{ab}) = \pm \Theta_{ab}$. So one has to integrate the dihedral angle with respect to the area. The result is known to be the Regge action, since when varying it with respect to $A_{ab}$, the variations of the dihedral angles cancel thanks to the Schlaefli identity, $\sum_{a<b} A_{ab}\,\delta\Theta_{ab} = 0$ \footnote{Strictly speaking, the Schlaefli identity is usually considered for any variations of the edge lengths. In the geometric sector, the set of spins and coherent states enables to reconstruct these lengths. Thus, the chain rule allows to write the Schlaefli identity for variations of the areas also.}.


To better control the approximation made in \eqref{semiclass eq}, it is useful to derive it directly from the asymptotic analysis of the models. The vertex amplitude $W$ is exactly defined as:
\beq
W = \int d\mu(X)\ \exp\Bigl(-\sum_{c<d} A_{cd} s_{cd}(X)\Bigr).
\ee
$X$ denotes the set of all variables to be integrated, and the function $s_{ab}$ has a positive real part. Thus, it can be evaluated using a saddle point and stationary phase method when rescaling the spins by a large $\lambda$. Applying the operator $\Delta_{ab}$ gives without approximation:
\beq
\Delta_{ab}\,W = 2\int d\mu(X)\, e^{-\sum A_{cd} s_{cd}(X)}\,\Bigl(\cos \bigl(s_{ab}(X)\bigr) -1\Bigr).
\ee
In the large spin limit, we can just evaluate $\cos(s_{ab}(X))$ on each saddle point. If it is the same for all of them, then it can be factorized from the amplitude. This is what happens in the $\SU(2)$ Ooguri model and the Lorentzian EPRL model, where the saddle points $X^*$ gives: $s_{ab}(X^*) = \pm \Theta_{ab}$. This point is important and prevents the asymptotics from getting other frequencies than the Regge action itself. As an example, our reasoning does not apply to the geometric sector of the Euclidean EPRL model, since it receives oscillations from the Regge action and from $\gamma\mone S_{\rm R}$ also, with the same scaling. Then, the spin foam amplitude satisfies a higher order difference equation in the asymptotics, which is simply the product of the difference operator \eqref{semiclass eq} for both frequencies:
\beq
\Bigl[\Delta_{ab} +2\,\bigl(1-\cos \f{\Theta_{ab}}{\gamma}\bigr)\Bigr]\,\Bigl[\Delta_{ab} +2\,\bigl(1-\cos \Theta_{ab}\bigr)\Bigr] V(A_{ab})= 0.
\ee

\subsection{Recursion relations as Wheeler-de-Witt equations in quantum area Regge calculus}

Since semi-classical spin foams can be approximated with quantum Regge calculus, we now want to understand our main equation \eqref{semiclass eq} in this framework. Here comes our second important point: this equation \eqref{semiclass eq} has a nice geometric interpretation as a quantization of a constraint in Regge calculus. Consider the set of areas $(A_{ab})$ such that they uniquely determine a genuine flat 4-simplex as the configuration space. Like in \cite{dittrich-ryan-simplicial-phase}, we take the conjugated momenta to be angles $(\theta_{ab})$, with the canonical brackets:
\beq
\{ A_{ab}, \theta_{cd}\} = \delta_{(ab),(cd)}.
\ee
This framework has been derived from a canonical discretization of the Plebanski's action for gravity in \cite{dittrich-ryan-simplicial-phase} (see there for the full details on the symplectic structure of the phase space). On this phase space, we also consider the constraints:
\beq \label{flat simplex}
\chi_{ab} \,\equiv\, \cos \theta_{ab} \,-\, \cos \Theta_{ab}(A) \,=\,0.
\ee
This was already studied in the above reference, and there argued to form an Abelian algebra. It should be noted that the authors of \cite{dittrich-ryan-simplicial-phase} were then interested in the gauge symmetry corresponding to the translation of a vertex of the simplex. Here, we would like instead to generate independent shifts of areas to produce our equation of interest \eqref{semiclass eq}. For that purpose, the constraint \eqref{flat simplex} is what we need. Furthermore, its geometric meaning is quite clear: the momenta $(\theta_{ab})$ are constrained to be the dihedral angles $(\Theta_{ab})$ of the flat 4-simplex determined by its areas.

Let us now proceed to the most naive quantization, using wave functions of the angles. They can be expanded onto the Fourier components, $(e^{i \sum j_{ab} \theta_{ab}})$, where the integers $(j_{ab})$ are the eigenvalues of the area operators $\hat{A}_{ab}$ (they get discrete spectra since the variables $\theta_{ab}$ live on a compact set). Then, periodic functions over $(\theta_{ab})$ act by multiplication, and in particular:
\begin{align}
\widehat{e^{\pm i\theta_{ab}}}\,\Bigl[\sum_{\{j_{cd}\}} \psi(j_{cd})\,e^{i\sum j_{cd}\theta_{cd}} \Bigr]
&= \sum_{\{j_{cd}\}} \psi(j_{ab}\mp 1, j_{cd})\,e^{i\sum j_{cd}\theta_{cd}},\\
&= \delta^\mp_{ab}\ \psi.
\end{align}
This simply means that on the Fourier coefficients $\psi(j_{cd})$ of a state $\vert \psi\rangle$, the operator $\widehat{e^{\pm i\theta_{ab}}}$ acts by shifting the variable $j_{ab}$ by $\mp 1$. Also, as the Fourier exponentials are the eigenfunctions of the area operators, we simply promote the complicated functions $\Theta_{ab}(A)$ to operators through:
\beq
\widehat{\cos\Theta_{ab}(A)}\, e^{i\sum j_{cd}\theta_{cd}} = \cos\Theta_{ab}(j)\,e^{i\sum j_{cd}\theta_{cd}},
\ee
as far as the set $(j_{cd})$ allows to define the dihedral angles. Thus the classical constraint $\chi_{ab}$ can be imposed at the quantum level,
\beq
\widehat{\chi_{ab}}\,\vert \psi \rangle \,=\,0,
\ee
where it becomes exactly the difference equation we are looking for:
\beq
\Bigl[\Delta_{ab} +2\,\bigl(1-\cos \Theta_{ab}\bigr)\Bigr] \psi(j_{ab}) = 0.
\ee

\section{Quantum dynamics of the flat 4-simplex in Loop quantum gravity \\(the Ooguri model revisited)} \label{sec:4d-wdw}

The above approach has obvious limitations:
\begin{itemize}
 \item First, it only holds asymptotically, and we expect the full quantum gravity amplitude to satisfy a difference equation with non-trivial coefficients, which would contain all information about the full asymptotic expansion.
 \item Second, it is not clear what the role of the additional boundary data of spin foams (the normals to the triangles) can be here. However, they are part of the phase space of Loop Quantum Gravity (on a single graph). In addition, it has been shown in previous studies that spin foams are better understood in terms of area-angle Regge calculus, \cite{bf-aarc-val}, instead of area calculus (well-known to suffer from several drawbacks), \cite{dittrich-speziale-aarc}. This leads to the third point.
 \item We have so far focused only on Regge geometries, since the asymptotic behaviour is different on the other configurations. But from the LQG point of view, there is no specific reason to distinguish between Regge and non-Regge geometries. Furthermore, the Ooguri model, which also shows up these different asymptotic behaviours, is nevertheless built from a single constraint, namely the flatness of a gauge field, without regards for the amount of geometricity contained in the canonical momenta.
\end{itemize}

This leads us to revisit the Ooguri model for $\SU(2)$ BF theory, with a new form, more geometric, of the Hamiltonian constraint.

\subsection{The proposal: projecting the curvature} \label{sec:proposal}

So we now turn to the phase space inherited from LQG on the dual complex $\Gamma$ to the boundary of a 4-simplex. The notation $a=1,\dotsc,5$ is kept for tetrahedra of the triangulation, and hence $(ab)$ for triangles and $(abc)$ for edges. By duality, they correspond on the complex $\Gamma$ to nodes $a=1,\dotsc,5$, links $(ab)$, and faces, also referred to as cycles $(abc)$. We will mainly use the terminology corresponding to cells of $\Gamma$, but also switch to the point of view of the triangulation as soon as we find it relevant.

The phase space is precisely the same as that of $\SU(2)$ Yang-Mills on such discretization, and the same as in the topological $\SU(2)$ BF theory, up to a scaling of the fundamental brackets by the Immirzi parameter (that we will ignore in the following). The ten links of $\Gamma$ carry $\SU(2)$ elements, $(g_{ab})$, which represent parallel transport operators between the source and target vertices of each link \footnote{We choose the notation so that $g_{ab}$ goes from $b$ to $a$. Also: $g_{ba} = g\mone_{ab}$.}. The phase space is then simply the cotangent bundle over $\SU(2)^L$, with $L=10$ and with its natural symplectic structure. The momentum to $g_{ab}$ is thus a 3-vector $E_{ab}$, for $a<b$, with
\beq
\{E^i_{ab}, g_{ab}\} = \tau^i\,g_{ab}.
\ee
These momenta can be seen as smearings the triad field $E^\alpha_i$ of the continuum over the triangles (dual to the link of $\Gamma$), and will be therefore called triad variables. We have denoted $(\tau^i)_{i=1,2,3}$ anti-hermitian generators of the Lie algebra. The standard interpretation is the following. Each tetrahedra of the boundary carries a local reference frame, and $E_{ab}$ is defined relatively to that of the tetrahedron $a$. Momenta $E_{ba}$ acting on the right of $g_{ab}$ can be defined by parallelly transporting $E_{ab}$ to the frame of the tetrahedron $b$ using the adjoint action of the group:
\beq \label{transport}
E_{ba} = - \Ad(g\mone_{ab})\,E_{ab}.
\ee
Like in lattice gauge theory, the gauge group is $\SU(2)^V$, where $V=5$ is the number of nodes of the graph.

We now consider a Hamiltonian constraint for the Ooguri model. It is usually taken to be:
\beq \label{trivial hol}
g_{(abc)} \equiv g_{ab} g_{bc} g_{ca} = \unit,
\ee
That is to say: the parallel transport around the cycle $(abc)$ made of the three links $(ab), (bc), (ca)$ is trivial. But it is unlikely that its quantization will help to understand quantum gravity. So we would like a constraint which would look like closer to the Hamiltonian constraint of general relativity, $\epsilon^{ij}_{\phantom{ij}k} E^\alpha_i E^\beta_j F(A)^k_{\alpha\beta} = 0$ (where $\alpha, \beta$ are indices on the canonical surface, and $A$ is a $\SU(2)$ gauge field of curvature $F$ and conjugated momentum $E$). First the curvature is discretized around the 2d regions of the complex dual to the triangulation. More precisely, consider the region bounded by the links $(ab), (bc), (ca)$, then the component of $F$ along these directions is regularized in LQG as:
\beq
\epsilon^{ij}_{\phantom{ij}k}\,F_{\alpha\beta}^k \quad \longrightarrow\quad \delta^{ij} - \bigl(\Ad(g_{(abc)})\bigr)^{ij}.
\ee
Then, the idea is to project it at each node of the cycle along the two triad variables which meet there. Define:
\beq
H^a_{bc} = E_{ab}\cdot E_{ac} - E_{ab}\cdot\Ad(g_{(abc)}) E_{ac},
\ee
and the constraint $H^a_{bc} = 0$. For each cycle, there are three such constraints (and for a generic triangulation, there is one constraint for each node of a cycle). So there are enough constraints to enforce $g_{(abc)}=\pm \unit$ when the triad variables around the cycle span the three dimensions \footnote{At least, it is not hard to see that there is no smooth deformation of this relation. However, there may be a finite set of possibilities that we have not investigated. In particular, the solutions of the equation $E_1\cdot E_2 - E_1\cdot \Ad(g) E_2=0$ are: $g= \exp(t_1 E_1)\exp(\eta (E_1\times E_2)) \exp(t_2 E_2)$, where $t_1, t_2$ are arbitrary. But $\eta$ admits only a finite number of values, since there is a finite number of $\SO(3)$ rotations with axis $E_1\times E_2$ solving the equation.}.

\subsection{A Hamiltonian for twisted geometries} \label{sec:twisted H}

We are interested in the relation between $H^a_{bc}$ and the previous constraint $\chi_{ab}$, \eqref{flat simplex}, at the classical level, and with the semi-classical equation \eqref{semiclass eq} after quantization. Classically, it is possible to define in the geometric sector the dihedral angles $\Theta_{ab}$ via the triad variables, and like in \cite{dittrich-ryan-simplicial-phase}, another notion of dihedral angles involving the group elements. Then, the above constraint just states the equality of the two notions. Details will appear in a collaboration with L. Freidel \cite{3d-wdw}.

Here, we prefer to translate the constraint into the language of twisted geometry \cite{freidel-speziale-twisted-geom}. This is a nice reparametrization of the LQG phase space which makes clear the nature of the involved geometries. In particular, space is formed by genuine polyhedra like tetrahedra, but their gluing does not lead to Regge metrics, since two adjacent polyhedra may describe their common boundary with different shapes. We take advantage of this fact and give an interpretation of $H^a_{bc}$ which also holds for non-Regge situations.

The parametrization maps the set $(E_{ab}, g_{ab})$ to a new set:
\beq
(E_{ab}, g_{ab})\quad \rightarrow \quad (A_{ab}, \vec{N}_{ab}, \vec{N}_{ba}, \xi_{ab}),
\ee
defined as follows. $A_{ab}$ is the norm of $E_{ab}$ (and equals that of $E_{ba}$), and $\vec{N}_{ab}$ its direction:
\beq
E_{ab} = A_{ab}\,\vec{N}_{ab},
\ee
for all $a,b$. Then, since $\vec{N}_{ab}$ and $\vec{N}_{ba}$ are taken as independent, the equation \eqref{transport}: $\vec{N}_{ab} = -\Ad(g_{ab}) \vec{N}_{ba}$ has to be solved for $g_{ab}$. Take a set of $\SU(2)$ rotations $(n_{ab}(\vec{N}))_{a,b}$ such that $n_{ab}$ maps an axis of reference in $\R^3$, say $\hat{z}$, onto $\vec{N}_{ab}$. This leads to the introduction of the angles $\xi_{ab}$ through:
\beq \label{twisted hol}
g_{ab} = n_{ab}\,\epsilon\  e^{\xi_{ab}\tau_z}\  n_{ba}\mone.
\ee
(The matrix $\epsilon=(\begin{smallmatrix} 0 &1\\-1 & 0\end{smallmatrix})$ is there to account for the minus sign in the parallel transport relation \eqref{transport}, since $\epsilon$ maps the direction $\hat{z}$ onto its opposite $-\hat{z}$.) Obviously, the normals and the triad variables are unchanged when adding a phase on the right of $n_{ab}$ like:
\beq \label{phase change}
n_{ab} \ \rightarrow\ n_{ab}\,e^{\lambda_{ab}\tau_z}.
\ee
The invariance of $g_{ab}$ then requires to change $\xi_{ab}$ accordingly. In particular, for a given $g_{ab}$, $\xi_{ab}$ can always be reabsorbed into the rotation $n_{ab}$ or $n_{ba}$. The set of rotations $(n_{ab})$ is very convenient, as we will see, and its use prefigures what happens at the quantum level. Indeed, the semi-classical coherent states we will later use are actually labelled by such rotations rather than only by the normals $(\vec{N}_{ab})$.

The generator of gauge transformations on the vertex dual to the tetrahedron $a$ is:
\beq \label{closure}
\sum_{b\neq a} A_{ab}\, \vec{N}_{ab} = 0,
\ee
This condition actually takes the form of a closure relation for the tetrahedron and hence leads to this nice interpretation: the variable $A_{ab}$ is the area of the triangle $(ab)$, while $\vec{N}_{ab}$ and $\vec{N}_{ba}$ are respectively the normals to the same triangle with respect to the frame of the tetrahedra $a$ and $b$. So it guarantees that one can built a flat tetrahedron in $\R^3$ for each $a$.

The areas and normals describe the intrinsic geometry of the canonical surface. In particular, in the gauge invariant sector
, the dihedral angle $\phi^a_{bc}$ between the triangles $(ab), (ac)$ is given by:
\beq \label{cos phi}
\cos\phi^a_{bc} =  - \frac{E_{ab}\cdot E_{ac}}{A_{ab}\,A_{ac}}.
\ee
Another key quantity we will need, defined only in terms of the six normals around a cycle $(abc)$, is:
\beq \label{4d angle}
\cos\Theta_{bc}^{(a)}(\vec{N}) = \frac{\cos\phi^a_{bc} - \cos\phi^b_{ac}\,\cos\phi^c_{ba}}{\sin\phi^b_{ac}\,\sin\phi^c_{ba}}.
\ee
If this quantity is independent of $a$ (that is computing it from any cycle containing the link $(bc)$ gives the same answer), then it is exactly the 4d angle $\Theta_{bc}$ between the tetrahedra $b$ and $c$, computed from the normals. This is exactly the criterion that turns a set of variables satisfying \eqref{closure} into a Regge metric on the triangulation \cite{dittrich-speziale-aarc}:
\beq \label{regge gluing}
\cos \Theta^{(a)}_{bc}(\vec{N}) =  \cos \Theta^{(a')}_{bc}(\vec{N}).
\ee

The splitting between intrinsic and extrinsic geometries in the twisted parametrization, and in particular, the information about the extrinsic geometry contained in the set of normals $(\vec{N}_{ab})$ has been discussed in \cite{freidel-speziale-twisted-geom}. Here, we are able to go further on this issue by writing the constraint $H^a_{bc}$ in this new set of variables. To get a definite expression, we need the Euler decomposition of the product $n_{ab}^{-1} n_{ac}$:
\beq \label{def alpha}
n_{ab}^{-1} n_{ac}^{\phantom{-1}} = e^{\alpha^a_{bc}\tau_z}\,e^{(\pi-\phi^a_{bc})\tau_y}\,e^{\alpha^a_{cb}\tau_z},
\ee
which defines the angles $(\alpha^a_{bc})_{a,b,c}$ (notice however that they are changed under \eqref{phase change}).

The Hamiltonian $H^a_{bc}$ then admits the following form on twisted geometries (see appendix):
\beq \label{twisted H}
H^a_{bc} = - A_{ab} A_{ac}\,\Bigl(\cos\phi^a_{bc} - \cos\phi^b_{ac}\,\cos\phi^c_{ba} \\
+ \sin\phi^b_{ac}\,\sin\phi^c_{ba}\,\cos\bigl(\xi_{bc} + \alpha^b_{ca} +\alpha^c_{ba}\bigr)\Bigr).
\ee
The surprise is that this is the form of the standard relation between the 3d and 4d dihedral angles within a flat 4-simplex, \eqref{4d angle}, though it holds on the whole phase space. Now restrict attention to the Regge-geometric sector (where \eqref{regge gluing} is satisfied). As soon as the tetrahedra are non-degenerate (the 3d angles are neither $0$ nor $\pi$), the angle $(\xi_{bc} + \alpha^b_{ca} +\alpha^c_{ba})$ can be extracted from the constraint $H^a_{bc}=0$, to give the dihedral angle $\Theta_{bc}$. Moreover, since $\xi_{bc}$ is also independent of $a$ obviously, then $(\alpha^b_{ca} +\alpha^c_{ba})$ also is. So we can use \eqref{phase change} to achieve a phase choice where this combination is $\pi$. This gives
\beq \label{geom flatness}
H^a_{bc} \propto \cos\Theta_{bc}(\vec{N}) \,-\, \cos\bigl(\xi_{bc} \bigr) = 0.
\ee
So the classical constraint $H^a_{bc} = 0$ really corresponds to the constraint $\chi_{ab}=0$ in their common domain of applicability (the geometric sector), and is clearly related to building a flat 4-simplex out of its boundary tetrahedra. A key difference is that the dihedral angles of the 4-simplex are rather computed from the set of normals $(\vec{N}_{ab})$ rather than from the areas. This suggests that the semi-classical limit of LQG is given by quantum area-angle Regge calculus. Further, it makes the formula applicable on the whole phase space, by going back to \eqref{twisted H} for $H^a_{bc}$, since the latter still makes sense outside of the geometric sector (when the gluing of the tetrahedra is not that of a 4-simplex), and also for degenerate tetrahedra (when some 3d angles are such that $\sin \phi^a_{bc}=0$).

We would like to mention that at this stage it is possible to discuss the solutions of the constraints \eqref{twisted H} in a way which is fully parallel to the analysis of \cite{barrett-asymEPR}. First, in the geometric sector, there are clearly two solutions,
\beq
\xi_{bc}=\pm \Theta_{bc}(\vec{N}).
\ee
Then, assume there are at least two distinct solutions, $(\xi^+_{bc}, \xi^-_{bc})$. That leads to: $E_{ba}\cdot\Ad(g^+_{bc}) E_{ca} = E_{ba}\cdot\Ad(g^-_{bc}) E_{ca}$. This relation has been studied in \cite{dittrich-ryan-simplicial-phase} where it was coined edge-simplicity constraint, and shown to actually imply the Regge gluing relations \eqref{regge gluing}. So in the non-geometric sector, the constraint has either one solution, or no solution.

\subsection{The Wheeler-de-Witt equation and its semi-classical regime}

\subsubsection{The Wheeler-de-Witt equation as recursion relations}

When a theory has gauge symmetries, the latter turn into constraints in the Hamiltonian analysis. They can be imposed either before or after quantizing. In general relativity and BF theory, the Hamiltonian itself is a constraint, and the program of LQG is to quantize first and then constrain. The kinematical Hilbert space is $\calH_\Gamma = L^2(\SU(2)^L/\SU(2)^V)$, spanned by the so-called spin network functions. These are just built from the Fourier modes of the ten group elements, that are their matrix elements in the representation $(j_{ab})_{a<b}$, while the magnetic numbers are contracted with a specific tensor $(\iota_a)$ on each node $a$, called intertwiner, which ensures gauge invariance.
\beq
s^{\{j_{ab},\iota_a\}}(g_{ab}) = \sum_{\{m_{ab}\}} \prod_{a<b} \langle j_{ab}\,m_{ab}\vert g_{ab}\vert j_{ab}\,m_{ba}\rangle \prod_a \iota_a^{\{m_{ab}\}}.
\ee
The quantization of operators is direct. Gauge invariant functions of the group elements act by multiplication, while $E^i_{ab}$ acts as a left derivative:
\beq
\widehat{E}^i_{ab}\,\langle j_{ab}\,m_{ab}\vert g_{ab}\vert j_{ab}\,m_{ba}\rangle = i\langle j_{ab}\,m_{ab}\vert \tau^i\,g_{ab}\vert j_{ab}\,m_{ba}\rangle.
\ee
In particular, the square of the above equation produces the area spectrum, that of the Casimir of $\SU(2)$.

The basic building block attached to every 4-simplex in Ooguri's model for $\SU(2)$ BF theory is a Wigner 15j-symbol. The latter can also be seen in the Hamiltonian framework as the flat state on the boundary of a 4-simplex, satisfying \eqref{trivial hol} on each cycle, expanded in the spin network basis.

Though it is not hard to express the action of the operator $E_{ab}\cdot E_{ac}$ on a spin network state, the result actually depends on the choice of a basis of intertwiner. A standard basis is obtained at each node of the graph by pairing the links meeting there, expanding the tensor products of their representations into irreducible representations, and choosing a spin common to the two tensor products. If $(ab)$ and $(ac)$ are paired together, then the operator $E_{ab}\cdot E_{ac}$ is diagonal on spin network states, \cite{baez-barrett-quantum-tet}. As for the operator $E_{ab}\cdot \Ad(g_{(abc)}) E_{ac}$, one first rewrite it as:
\beq
E_{ab}\cdot \Ad(g_{(abc)}) E_{ac} = E_{ba}\cdot \Ad(g_{bc}) E_{ca}.
\ee
The triad variables produce insertions of generators on the link $(ab)$ at the node $b$, and on the link $(ac)$ at the node $c$. The action is not diagonal since $\Ad(g_{bc})$ is a multiplication by matrix elements in the spin 1 representation. This produces shifts on the spin $j_{bc}$ to $j_{bc}+\kappa$, for $\kappa = -1,0,1$.

Depending on the chosen pairings on the spin network nodes, the action of $H^a_{bc}$ thus leads to different equations (all of them being difference equations of the second order acting on one or more spins). However, using the same tools and methods as those of \cite{3d-wdw}, it can be oberved that these difference equations exactly take the form of recursion relations satisfied by the Wigner $\SU(2)$ 15j-symbol \footnote{The generic process is that the action of a triad variable inserts a generator, and then, the contraction of their vector indices produce graspings on the spin network in the spin 1 representation. After some recoupling, one can extract a special 6j-symbol with a spin 1 at each node of the cycle (including the virtual spins of intertwiners).}. On the one hand, this 15j-symbol is the basic building block attached to every 4-simplex in Ooguri's model for $\SU(2)$ BF theory. On the other hand, it is in the Hamiltonian framework the flat state on the boundary of a 4-simplex, satisfying \eqref{trivial hol} on each cycle, when expanded in the spin network basis. The recursion relations have been derived in \cite{recurrence-paper}, from the topological invariance spin foam model -- a method which already suggested a strong connection to the classical symmetry of the theory.

Assume for instance that the intertwiner $\iota_1$ pairs $(j_{12}, j_{13})$ together to a virtual spin $i_1$, that $\iota_2$ pairs $(j_{12},j_{23})$ to $i_2$ and $\iota_3$ pairs $(j_{23},j_{13})$ to a virtual spin $i_3$. Then, the Wheeler-de-Witt equation:
\beq
\widehat{H}^1_{23}\ \vert\psi \rangle = 0,
\ee
is actually the same as the recursion relation on the 6j-symbol,
\beq
A_{-1}(j_{12})\,\psi(j_{12}-1) + A_0(j_{12})\,\psi(j_{12}) + A_{+1}(j_{12})\,\psi(j_{12}+1) \,=\,0.
\ee
The coefficients $A_{\pm 1}$ take the form: $A_{+1}(j) = j E(j+1)$, and $A_{-1}(j) = (j-1) E(j)$, for
\be
E(j_{12}) = \Bigr[\bigl((j_{13}+i_1+1)^2-j_{12}^2\bigr) \bigl(j_1^2-(j_{13}-i_1)^2\bigr) \bigl((j_{23}+i_2+1)^2-j_{12}^2\bigr) \bigl(j_{12}^2-(j_{23}-i_2)^2\bigr)\Bigr]^{\f12},
\ee
and the coefficient $A_0$ is given by:
\begin{multline}
A_0(j_{12}) = \bigl(2j_{12}+1\bigr)\Bigl\{2\bigl[j_{13}(j_{13}+1)i_2(i_2+1)+j_{23}(j_{23}+1)i_1(i_1+1)-j_{12}(j_{12}+1)i_3(i_3+1)\bigr] \\
- \bigl[j_{13}(j_{13}+1)+i_1(i_1+1)-j_{12}(j_{12}+1)\bigr]\bigl[j_{23}(j_{23}+1)+i_2(i_2+1)-j_{12}(j_{12}+1)\bigr]\Bigr\}.
\end{multline}

If now on the node $a=1$, the intertwiner $\iota_a$ pairs $(j_{12}, j_{15})$ together to the virtual spin $i_1$, then the equation becomes more complicated:
\beq
\sum_{\epsilon_{12}, \epsilon_1 = -1,0,1} A_{\epsilon_{12}, \epsilon_1}(j_{12}, i_1)\, \psi(j_{12}+\epsilon_{12}, i_1+\epsilon_1) = 0.
\ee
Such relations were derived by writing down explicitly a special invariance of the Ooguri model under a change of triangulation. Thus, it was known that these relations encode the symmetries of the model at the quantum level. But they had so far never been derived from the quantization of a Hamiltonian constraint.

\subsubsection{The Wheeler-de-Witt equation in the semi-classical regime}

The above equation is not really suitable for the semi-classical analysis however. It comes from the uncertainty principle, that a tetrahedron is described quantum mechanically by only five quantum numbers (four areas $(j_{ab})$, and one spin $i_a$ to specify the intertwiner) \cite{baez-barrett-quantum-tet}. So to launch our Wheeler-de-Witt equation in the semi-classical limit, we go to an overcomplete basis of coherent intertwiners \cite{livine-coherentBF}. First build the usual $\SU(2)$ coherent state $\vert j, n(\vec{N})\rangle$ from a $\SU(2)$ rotation $n(\vec{N})$ which maps the reference axis $\hat{z}$ onto a unit 3-vector $\vec{N}$:
\beq
\vert j, n(\vec{N})\rangle = n(\vec{N})\,\vert j, j\rangle.
\ee
It is important to keep in mind that the state is not fully determined by the direction $\vec{N}$, but also by a choice of phase. Indeed, changing $n$ like in \eqref{phase change} does not affect the vector $\vec{N}$, but multiplies the state by a phase. A coherent intertwiner $\iota_a(n_{ab})$ on the tetrahedron $a$ is labelled by four rotations $(n_{ab})_{b\neq a}$ corresponding to four unit vectors of $\R^3$, $(\vec{N}_{ab})_{b\neq a}$. It is defined by a group averaging process:
\beq \label{coherent intertwiner}
\int_{\SU(2)} dh_a\ \otimes_{b\neq a} h_a\,\vert j_{ab}, n_{ab}\rangle.
\ee
It is shown in \cite{livine-coherentBF} that the norm of this intertwiner is peaked for large spins on vectors $(\vec{N}_{ab})$ which satisfy the closure condition \eqref{closure}, but with quantum areas $(j_{ab})$. Therefore, these vectors can be interpreted as normals to the triangles of $a$ (up to a global rotation). The spin network state in the basis of coherent intertwiners reads:
\beq
s^{\{j_{ab}, n_{ab}\}}(g_{ab} ) = \int_{\SU(2)^5} \prod_{a=1}^5 dh_a\ \prod_{a<b} \langle j_{ab}, n_{ab}\epsilon \vert\, h_a\mone\,g_{ab}\,h_b\,\vert j_{ab}, n_{ba}\rangle.
\ee
(The matrix $n_{ab}\epsilon$ sends the axis of reference $\hat{z}$ onto $-\vec{N}_{ab}$. This choice is a matter of convenience.)

To evaluate the action of $H^a_{bc}$ on such states, some work is necessary, which is reported in appendix. However, the result is very natural. Indeed, the data of these coherent states match the data $(A_{ab}, \vec{N}_{ab},\vec{N}_{ba})$ of twisted geometries. So it is expected, if they really have a nice semi-classical behaviour, that they will lead to some simple quantum version of the classical Hamiltonian on twisted geometries \eqref{twisted H}.

The nice semi-classical behaviour is inherited from the following property of $\SU(2)$ coherent states:
\beq 
\vec{\tau}\, \vert j,n(\vec{N})\rangle = (-i)\,j\vec{N} \,\vert j,n(\vec{N})\rangle + o(j).
\ee
Since the triad operator $E_{ab}$ inserts a generator like this, we can guess the action of $E_{ab}\cdot E_{ac}$:
\beq
\Bigl(\widehat{E_{ab}\cdot E_{ac}}\Bigr)\,s^{\{j_{ab},n_{ab}\}} \simeq j_{ab} j_{ac}\ \cos\phi^a_{bc}\ s^{\{j_{ab},n_{ab}\}}.
\ee
Acting with $E_{ba}\cdot\Ad(g_{bc}) E_{ca}$ is a bit more involved, but there are no conceptual difficulties. The adjoint action of $g_{bc}$ is a Wigner matrix in the representation of spin 1, which recouples with the matrix elements of $g_{bc}$ in the state. This produces some shifts of the spin $j_{bc}$ to $j_{bc}+\eta$, for $\eta=-1,0,1$. These shifts extend to the coherent states, to $\vert j_{bc}+\eta,n_{bc}\rangle, \langle j_{bc}+\eta, n_{cb}\epsilon\vert$ in the large spin limit, via the operators $E_{ba}, E_{ca}$. We thus get:
\beq
\Bigl(\widehat{E_{ba}\cdot\Ad(g_{bc}) E_{ca}}\Bigr) s^{\{j_{ab},n_{ab}\}} \simeq \cos\phi^b_{ac}\,\cos\phi^c_{ba}\ s^{j_{bc}} - \f12 \sin\phi^b_{ac}\,\sin\phi^c_{ba}\Bigl( e^{-i(\alpha^b_{ca} +\alpha^c_{ba})}\, \delta^+_{bc} + e^{i(\alpha^b_{ca} +\alpha^c_{ba})}\, \delta^-_{bc}\Bigr)\, s^{j_{bc}},
\ee
where the angles $\alpha$ are determined by:
\beq
n_{ab}^{-1} n_{ac}^{\phantom{-1}} = e^{\alpha^a_{bc}\tau_z}\,e^{(\pi-\phi^a_{bc})\tau_y}\,e^{\alpha^a_{cb}\tau_z},
\ee
like for classical twisted geometries.

The final step is to rewrite the quantum condition:
\beq
\widehat{H}^a_{bc}\ \vert \psi\rangle =0,
\ee
for the coefficients of an arbitrary expansion in the coherent spin network basis. Again, we have to invoke the semi-classical, large spin limit, since coherent intertwiners generally have non-trivial overlap. They become orthogonal for large spins, and then the Wheeler-de-Witt equation for $H^a_{bc}$ reads:
\beq \label{4d-wdw}
\Bigl(\cos\phi^a_{bc} - \cos\phi^b_{ac}\,\cos\phi^c_{ba}\Bigr)\,\psi(j_{bc}) + \f12 \sin\phi^b_{ac}\,\sin\phi^c_{ba}\Bigl(e^{i(\alpha^b_{ca} +\alpha^c_{ba})} \psi(j_{bc}+1) + e^{-i(\alpha^b_{ca} +\alpha^c_{ba})} \psi(j_{bc}-1)\Bigr) = 0.
\ee
Here the dependence on other variables than $j_{bc}$ have been dropped. This is our key equation, which generalizes the semi-classical equation satisfied by the exponential of the Regge action \eqref{semiclass eq} to the whole phase space of twisted geometries (with sufficiently large spins).

The following is devoted to specializing \eqref{4d-wdw} to Regge or non-Regge boundary data, and see that it reproduces the equations and results previously discussed. We look for solving the equation {\it \`a la WKB}, when all spins are rescaled by $\lambda \gg 1$, and with the ansatz:
\beq
\psi(\lambda j_{bc}) = \Phi(j_{bc})\,e^{i S(\lambda j_{bc})}.
\ee
We assume $\Phi$ does not scale with $\lambda$, while $S$ grows linearly. 
To zeroth order, $\psi(\lambda j_{bc}\pm 1) \simeq \psi(\lambda j_{bc})\,e^{\pm iS'(\lambda j_{bc})}$,
where $S'$ is the derivative of $S$ seen as a function on the real line. So, the equation \eqref{4d-wdw} becomes:
\beq \label{solve wdw}
\Bigl(\cos\phi^a_{bc} - \cos\phi^b_{ac}\,\cos\phi^c_{ba}\Bigr) + \sin\phi^b_{ac}\,\sin\phi^c_{ba}\ \cos\bigl(S'(j_{bc}) + \alpha^b_{ca} +\alpha^c_{ba}\bigr) = 0.
\ee
We can now make contact with the first semi-classical equation of the paper \eqref{semiclass eq}, via the classical constraint $H^a_{bc}$ written for twisted geometries in \eqref{twisted H}. Indeed, the latter is the same as this, with $S'(j_{bc})$ instead of the angle $\xi_{bc}$.

Assume non-degeneracy of the tetrahedra, so that the formula \eqref{4d angle} for the 4d dihedral angles $\Theta_{ab}$ as functions of the normals $(\vec{N}_{ab})$ is well-defined. Further assume the boundary data satisfy the gluing constraints \eqref{regge gluing} and the closure relation \eqref{closure}. Then, we know from the discussion above \eqref{geom flatness} that: (i) the quantity into brackets on the left of \eqref{solve wdw} is: $\sin\phi^b_{ac}\,\sin\phi^c_{ba}\,\cos\Theta_{bc}$, (ii) the angle $(\alpha^b_{ca} +\alpha^c_{ba})$ can be set to $\pi$ by a change of phase in the coherent states $\vert j_{ab},n_{ab}(\vec{N})\rangle$. Then, we get to \eqref{solve semiclass},
\beq
\cos S'(j_{bc}) - \cos\Theta_{bc} = 0,
\ee
solved by exponentials of $\pm i$ times the Regge action $S_{\rm R}$. Now assume that \eqref{solve wdw} has only one solution for $S'$. The same way as it is discussed in \cite{barrett-asymEPR}, the value of $S'$ can be reabsorbed into a change of phase of the coherent states (this change of phase depends on the boundary data $(j_{ab}, \vec{N}_{ab}, \vec{N}_{ba})$). Such a choice cancels the oscillations on the wave function, $S=0$, like in the asymptotic analysis of \cite{barrett-asym15j}.

\section{Outlook} \label{sec:outlook}

The equation \eqref{semiclass eq} has been shown to come from a constraint stating that the momenta conjugated to the areas have to be the dihedral angles of a flat 4-simplex with these values of area (in the geometric sector). So this equation encodes the full information on the reconstructed 4-simplex, and we can imagine deriving from it other equations containing a few less geometric content, but still relevant. We here give one interesting example, which is a difference equation probing the closure of the simplex. Indeed, we can form the $5\times 5$ Gram matrix:
\beq
(G_{ab})\,=\,\bigl(\cos\Theta_{ab}\bigr).
\ee
Since the angles are those of a flat 4-simplex (actually determined by the areas and normals $(A_{ab}, \vec{N}_{ab}, \vec{N}_{ba})_{a<b}$, but this information will be lost in the coming equation), its determinant vanishes:
\beq \label{zero gram}
\det\,(G_{ab}) = 0.
\ee
From \eqref{semiclass eq}, we know that semi-classically, the multiplication by $\cos\Theta_{ab}$ can be compensated with shifts on the spin, by $(\delta^-_{ab}+\delta^+_{ab})$. This leads to the definition of a \emph{closure operator} which annihilates the 4-simplex spin foam amplitude $V(j_{ab})$ in the large spin limit:
\beq \label{closure op}
\det\ \Bigl(\f12\bigl(\delta^+_{ab} + \delta^-_{ab}\bigr)\Bigr)\ V(j_{ab}) \simeq 0.
\ee
This equation also holds for the Lorentzian EPRL model with shifts of the areas $\gamma j_{ab}$ instead. It can be directly checked, and this has been done in \cite{recurrence-paper}, that the exponential of $i$ times the Regge action is indeed a solution in the large spin limit, the key technical point being the use of \eqref{zero gram}.

It is remarkable that a 4-simplex spin foam amplitude satisfying this recursion relation in an exact way is actually known: the Barrett-Crane 10j-symbol (see \cite{recurrence-paper})\footnote{Note that the natural areas in the Barrett-Crane model are $(2j_{ab}+1)$, and not the spins themselves. So the operators $\delta^\pm_{ab}$ should then be understood as generating shifts of $\pm\f12$ in the spins (or equivalently, shifts of $\pm 1$ on the areas).}. Now that we understand that \eqref{closure op} is naturally a consequence of the flatness constraint (or semi-classical difference equation \eqref{semiclass eq}), this result for the 10j-symbol is even more interesting. Indeed, it is unlikely that the 10j-symbol is annihilated by the operator $[\f12(\delta^+ +\delta^-) -\cos\Theta]$, since the rapid oscillations due to the Regge action only appear in subleading orders in the asymptotics. Nevertheless, it turns out that it satisfies the weaker constraint corresponding to the closure.

The Hamiltonian $H^a_{bc}$ we defined for the Ooguri model can be seen as a discretization of a quantity, $\epsilon^{ij}_{\phantom{ij}k} E^\alpha_i E^\beta_j F^k_{\alpha\beta}$, for a fixed $\alpha$ and $\beta$ (that is a fixed face of the complex dual to the triangulation). So it is natural to try to implement the sum of the space indices to realize the Hamiltonian constraint of general relativity. A simple (naive ?) idea is to sum over the faces (cycles) that meet at the vertex $a$ on the boundary of a 3-cell, say $(abcd)$. (Viewed from the triangulation, it means summing over the edges which meet a node within a tetrahedron),
\beq
H^a_{bc} + H^a_{cd} + H^a_{db} =0.
\ee
From our analysis, it is clear that exponentials of $\pm i$ times the Regge action are annihilated in the large spin regime by this constraint (the three contributions would cancel independently). But we expect that there may be more solutions than for $H^a_{bc}=0$ which leads to the 15j-symbol at the quantum level. Since $H^a_{bc}$ generates in some way the symmetry and the moves at the core of the topological invariance of the Ooguri model, the above constraint would relax that to asking for an invariance under a combination of the moves. This idea was argued in \cite{recurrence-paper} on a toy example, but we may consider the above quantity as a more interesting realization.

Such an operator is labelled by a node and 3-cell (instead of a node and a cycle for $H^a_{bc}$). Note that in the case we have considered, that of the boundary of a 4-simplex, these 3-cells are actually of tetrahedral form. So pictorially speaking, this strongly echoes the recent proposal \cite{alesci-rovelli-hamiltonian} of regularization of the Hamiltonian constraint in LQG. The latter consists in adding a small tetrahedral graph at the node of a spin network grah. An important feature is that a 15j-symbol could be extracted there. So it is definitely a good direction to compare the simple approach presented here (which perfectly works for the topological Ooguri model) with the standard approach based on the Thiemann trick, and more specifically on the volume operator of LQG (the latter does not seem to play any role here, a priori).


\section*{Acknowledgements}
The author is indebted to Laurent Freidel and Etera Livine. Numerous ideas actually came from them.

\appendix
\section*{Appendix: Technical details}
\subsection*{On the Hamiltonian for twisted geometries}

Since the normals $\vec{N}_{ab}$ can be defined by:
\beq
\vec{N}_{ab} = \Ad(n_{ab})\,\hat{z},
\ee
we can easily use the rotations $(n_{ab})$ to capture the geometric information. Let us form:
\beq
n_{ab}^{-1} n_{ac}^{\phantom{-1}} = e^{\alpha^a_{bc}\tau_z}\,e^{(\pi-\phi^a_{bc})\tau_y}\,e^{\alpha^a_{cb}\tau_z}.
\ee
This is the Euler decomposition of the product $n_{ab}^{-1} n_{ac}^{\phantom{-1}}$. All that we need can be extracted from the matrix elements of this product in the representation of spin 1. Indeed, one can check that $\phi^a_{bc}$ as it appears above is the (3d) dihedral angle between $(ab), (ac)$:
\beq \label{dot product vectors}
\langle 1,0\vert\,n_{ab}^{-1} n_{ac}^{\phantom{-1}}\,\vert 1,0\rangle =  \frac{E_{ab}\cdot E_{ac}}{A_{ab}\,A_{ac}} = - \cos\phi^a_{bc} .
\ee
Furthermore, the sine of the dihedral angle can also be directly extracted:
\beq
\langle 1,\pm 1\vert \,n_{ab}^{-1} n_{ac}^{\phantom{-1}}\,\vert 1,0\rangle = \pm\f{1}{\sqrt{2}}\,e^{\mp i \alpha^a_{bc}}\ \sin\phi^a_{bc},\qquad 
\langle 1,0\vert \,n_{ab}^{-1} n_{ac}^{\phantom{-1}}\,\vert 1,\pm1 \rangle = \mp\f{1}{\sqrt{2}}\,e^{\mp i \alpha^a_{cb}}\ \sin\phi^a_{bc}.
\ee

Let us now prove the formula \eqref{twisted H}. Similarly to \eqref{dot product vectors}, one has:
\beq
E_{ba}\cdot \Ad(g_{bc}) E_{ca} = \langle 1,0\vert \ n_{ba}\mone\, g_{bc}\,n_{ca}\ \vert 1,0\rangle,
\ee
by definition of the different group elements here involved. Then, write $g_{bc}$ in terms of the variables of twisted geometries, \eqref{twisted hol}:
\beq
E_{ba}\cdot \Ad(g_{bc}) E_{ca} = \langle 1,0\vert \ (n_{ba}\mone\, n_{bc})\,\epsilon\,e^{\xi_{bc}\tau_z}\,(n_{cb}\mone\,n_{ca})\ \vert 1,0\rangle.
\ee
The matrix elements of $\epsilon = e^{-\pi\tau_y}$ in the representation of spin $j$ are: $\langle j,m\vert \epsilon\vert j,m'\rangle = (-1)^{j-m'}\delta_{m,-m'}$. Hence, introducing a resolution of the identity:
\beq
E_{ba}\cdot \Ad(g_{bc}) E_{ca} = \sum_{m=-1,0,1} \langle 1,0\vert \ n_{ba}\mone\, n_{bc}\ \vert 1,m\rangle\, (-1)^{1-m}\,e^{im\xi_{bc}}\,\langle 1,-m\vert\ n_{cb}\mone\,n_{ca}\ \vert 1,0\rangle.
\ee
So \eqref{twisted H} comes from explicitly writing down the three terms in the sum and the matrix elements.


\subsection*{Quantization of the Hamiltonian in the large spin limit}

To evaluate the action of $H^a_{bc}$ on coherent spin network states, let us first collect a few results. The action of the generators of $\su(2)$ is what we expect from states with a nice semi-classical behaviour:
\beq \label{generator action}
\vec{\tau}\, \vert j,n(\vec{N})\rangle = (-i)\,j\vec{N} \,\vert j,n(\vec{N})\rangle + o(j).
\ee
It is natural to see the 3-vector $\vec{N}$ in the Lie algebra $\su(2)$. Then, its components on the spherical basis $(\tau_m)_{m=-1,0,1}$ are:
\beq \label{N components}
N^m = \langle 1,m\vert n(\vec{N})\vert 1,0\rangle,\quad m=-1,0,+1.
\ee
$\SU(2)$ invariant operators acting on a coherent intertwiner naturally commute with the group action in \eqref{coherent intertwiner}.
In particular, these results can be used to compute the action of $E_{ab}\cdot E_{ac}$. Each triad operator inserts a generator, so that we have to consider the action of $(\sum_{i} \tau^i_{ab}\otimes \tau^i_{ac})$ on the coherent intertwiner. It commutes with the group action, and finally leads to:
\begin{align}
\Bigl(\widehat{E_{ab}\cdot E_{ac}}\Bigr)\,s^{\{j_{ab},n_{ab}\}} &\simeq  - j_{ab} j_{ac}\,\bigl(\vec{N}_{ab}\cdot \vec{N}_{ac}\bigr)\,s^{\{j_{ab},n_{ab}\}},\\
&\simeq \ j_{ab} j_{ac}\ \cos\phi^a_{bc}\ s^{\{j_{ab},n_{ab}\}}.
\end{align}
Acting with $E_{ba}\cdot\Ad(g_{bc}) E_{ca}$ is a bit more involved, but there are no conceptual difficulties. First the adjoint action of $g_{bc}$ is a Wigner matrix $D^{(1)}$ in the spin 1, which recouples to the matrix elements of $g_{bc}$ in the state, using Clebsch-Gordan coefficients:
\beq
D^{(j_{bc})}_{AB}(g_{bc})\ D^{(1)}_{kp}(g_{bc}) = \sum_{\eta=-1,0,1} C^{j_{bc}}_A{}^1_k{}^{j_{bc}+\eta}_{A+k}\ C^{j_{bc}}_B{}^1_p{}^{j_{bc}+\eta}_{B+p}\ D^{(j_{bc}+\eta)}_{A+k\,B+p}(g_{bc}).
\ee
The question is then whether (and since the answer is yes, how) the coherent state $\vert j_{bc},n_{bc}\rangle$ in the state also receives a shift by $\eta=-1,0,1$ consistently. This actually comes from the triad operators. Combining \eqref{generator action} and \eqref{N components}, the action of $E_{ba}$ is recast so that the quantity of interest is a state in the tensor product of the representation spaces $\calH_{j_{bc}}\otimes \calH_1$:
\beq \label{detail}
\vert j_{bc}, n_{bc}\rangle \otimes n_{ba}\vert 1,0\rangle = \sum_{m=-1,0,1} \Bigl(\langle 1,m\vert\, n_{bc}\mone n_{ba}\,\vert 1,0\rangle\Bigr)\ \vert j_{bc},n_{bc}\rangle \otimes n_{bc}\vert 1,m\rangle.
\ee
On the right hand side, we have just introduced the identity on $\calH_1$ as: $\sum_m n_{bc}\vert 1,m\rangle \langle 1,m\vert n_{bc}\mone$. The quantity into brackets can then be evaluated thanks to the formula \eqref{def alpha} which has already been used for twisted geometries. Quite clearly, this introduces the cosine of the dihedral angle $\phi^b_{ac}$ when $m=0$, and its sine, with some phase $\exp(\mp i\alpha^b_{ca})$ when $m=\pm 1$. Then, a key technical point which leads to the final expression is due to a careful inspection of the scaling properties of the Clebsch-gordan coefficients in $\calH_j\otimes \calH_1$ for large $j$:
\beq
\vert j, n(\vec{N})\rangle \otimes n(\vec{N})\vert 1, m\rangle \simeq \vert j+m, n(\vec{N})\rangle.
\ee
This is exact when $m=1$, but only holds asymptotically for $m=-1,0$. It enables to reintroduce a coherent state, with a shift on the spin, on the right hand side of \eqref{detail}:
\beq
\vert j_{bc}, n_{bc}\rangle \otimes n_{ba}\vert 1,0\rangle \simeq \Bigl[ -\cos\phi^b_{ac}\,\vert j_{bc},n_{bc}\rangle + \f{1}{\sqrt{2}}e^{-i\alpha^b_{ca}}\,\sin\phi^b_{ac}\,\vert j_{bc} +1,n_{bc}\rangle - \f{1}{\sqrt{2}}e^{i\alpha^b_{ca}}\,\sin\phi^b_{ac}\,\vert j_{bc} -1,n_{bc}\rangle\Bigr]
\ee
Obvisouly, a similar result holds for $\langle j_{bc}, n_{cb}\epsilon\vert$, on the other end of the link $(bc)$, upon contraction with $E_{ca}$. Gathering these different pieces, we get:
\beq
\Bigl(\widehat{E_{ba}\cdot\Ad(g_{bc}) E_{ca}}\Bigr) s^{\{j_{ab},n_{ab}\}} \simeq j_{ab} j_{ac}\biggl[\cos\phi^b_{ac}\,\cos\phi^c_{ba} s^{j_{bc}} - \f12 \sin\phi^b_{ac}\,\sin\phi^c_{ba}\Bigl( e^{-i(\alpha^b_{ca} +\alpha^c_{ba})} s^{j_{bc}+1} + e^{i(\alpha^b_{ca} +\alpha^c_{ba})} s^{j_{bc}-1}\Bigr)\biggr].
\ee




\begin{thebibliography}{10}

\bibliographystyle{amsplain}

\bibitem{alesci-rovelli-hamiltonian}
E.~Alesci and C.~Rovelli, \emph{{A regularization of the hamiltonian constraint
  compatible with the spinfoam dynamics}},  (2010).

\bibitem{baez-barrett-quantum-tet}
J.~C. Baez and J.~W. Barrett, \emph{{The quantum tetrahedron in 3 and 4
  dimensions}}, Adv. Theor. Math. Phys. \textbf{3} (1999), 815--850.

\bibitem{barrett-asymEPR}
J.~W. Barrett, R.~J. Dowdall, W.~J. Fairbairn, H.~Gomes, and F.~Hellmann,
  \emph{{Asymptotic analysis of the EPRL four-simplex amplitude}}, J. Math.
  Phys. \textbf{50} (2009), 112504.

\bibitem{barrett-asym-summary}
J.~W. Barrett, R.~J. Dowdall, W.~J. Fairbairn, H.~Gomes, F.~Hellmann, and
  R.~Pereira, \emph{{Asymptotics of 4d spin foam models}},  (2010).

\bibitem{pereira-asymEPR}
J.~W. Barrett, R.~J. Dowdall, W.~J. Fairbairn, F.~Hellmann, and R.~Pereira,
  \emph{{Lorentzian spin foam amplitudes: graphical calculus and asymptotics}},
   (2009).

\bibitem{barrett-asym15j}
J.~W. Barrett, W.~J. Fairbairn, and F.~Hellmann, \emph{{Quantum gravity
  asymptotics from the SU(2) 15j symbol}}, Int. J. Mod. Phys. \textbf{A25}
  (2010), 2897--2916.

\bibitem{bf-aarc-val}
V.~Bonzom, \emph{{From lattice BF gauge theory to area-angle Regge calculus}},
  Class. Quant. Grav. \textbf{26} (2009), 155020.

\bibitem{3d-wdw}
V.~Bonzom and L.~Freidel, \emph{{The Hamiltonian constraint in 3d Riemannian
  loop quantum gravity}}, To appear.

\bibitem{recurrence-paper}
V.~Bonzom, E.~R. Livine, and S.~Speziale, \emph{{Recurrence relations for spin
  foam vertices}}, Class. Quant. Grav. \textbf{27} (2010), 125002.

\bibitem{dittrich-ryan-simplicial-phase}
B.~Dittrich and J.~P. Ryan, \emph{{Phase space descriptions for simplicial 4d
  geometries}},  (2008).

\bibitem{dittrich-speziale-aarc}
B.~Dittrich and S.~Speziale, \emph{{Area-angle variables for general
  relativity}}, New J. Phys. \textbf{10} (2008), 083006.

\bibitem{maite-etera-6jcorr}
M.~Dupuis and E.~R. Livine, \emph{{The 6j-symbol: Recursion, Correlations and
  Asymptotics}}, Class. Quant. Grav. \textbf{27} (2010), 135003.

\bibitem{freidel-louapre-6j}
L.~Freidel and D.~Louapre, \emph{Asymptotics of 6j and 10j symbols},
  Class.Quant.Grav. \textbf{20} (2003), 1267--1294, [arXiv:hep-th/0209134].

\bibitem{freidel-speziale-twisted-geom}
L.~Freidel and S.~Speziale, \emph{{Twisted geometries: A geometric
  parametrisation of SU(2) phase space}},  (2010).

\bibitem{livine-coherentBF}
E.~R. Livine and S.~Speziale, \emph{{A new spinfoam vertex for quantum
  gravity}}, Phys. Rev. \textbf{D76} (2007), 084028.

\bibitem{ooguri4d}
H.~Ooguri, \emph{{Topological lattice models in four-dimensions}}, Mod. Phys.
  Lett. \textbf{A7} (1992), 2799--2810.

\bibitem{perez-intro-lqg}
A.~Perez, \emph{{Introduction to loop quantum gravity and spin foams}},
  (2004).

\bibitem{PR}
G.~Ponzano and T.~Regge, \emph{{Semi-classical limit of Racah coefficients}},
  Spectroscopic and group theoretical methods in physics (F. Bloch, ed.),
  North-Holland, Amsterdam (1968).

\bibitem{rovelli-PRTVO}
C.~Rovelli, \emph{{The Basis of the Ponzano-Regge-Turaev-Viro-Ooguri quantum
  gravity model in the loop representation basis}}, Phys. Rev. \textbf{D48}
  (1993), 2702--2707.

\bibitem{rovelli-new-look-lqg}
C.~Rovelli, \emph{{A new look at loop quantum gravity}},  (2010).

\bibitem{schulten-gordon2}
K.~Schulten and R.~G. Gordon, \emph{{Semiclassical approximations to 3J and 6J
  coefficients for quantum mechanical coupling of angular momenta}}, J. Math.
  Phys. \textbf{16} (1975), 1971--1988.

\end{thebibliography}


\end{document}